\documentclass[1p,preprint,10pt]{elsarticle}
\usepackage{amsmath}
\usepackage{amssymb}
\usepackage{amsmath}
\usepackage{textcomp}
\usepackage{color}
\usepackage{framed}
\usepackage{hyperref}
\hypersetup{colorlinks=true,breaklinks=true,linkcolor=blue,urlcolor=blue,citecolor=blue}
\definecolor{shadecolor}{rgb}{0.9,0.9,0.9}
\newcommand\Let{\mathrel{\mathop:\!\!=}}

\newcommand{\kv}{\ensuremath{\mathbf{k}}}

\def \min {\mathop {\rm min}}

\newcommand{\pdag}{{\phantom{\dagger}}}
\newcommand{\up}{\ensuremath{\uparrow}}
\newcommand{\down}{\ensuremath{\downarrow}}

\newcommand{\aver}[1]{\ensuremath{\left\langle{#1}\right\rangle}}
\newcommand{\abs}[1]{\ensuremath{\lvert#1\rvert}}

\journal{Computer Physics Communications}

\usepackage{graphicx}

\usepackage{amssymb}

\usepackage{lineno}

\definecolor{darkblue}{rgb}{0,0,.6}
\definecolor{darkred}{rgb}{.6,0,0}
\definecolor{darkgreen}{rgb}{0,.6,0}
\definecolor{red}{rgb}{.98,0,0}

\def\ssmall{\fontsize{8pt}{2pt}\selectfont}

\usepackage{listings}
\usepackage{caption}

\DeclareCaptionFormat{listing}{\rule{\dimexpr\textwidth+17pt\relax}{0.4pt}\par\vskip1pt#1#2#3}
\lstnewenvironment{cpplisting}[1][]
  {\lstset{language=C++,numbers=right,#1}}{}

\lstnewenvironment{pylisting}[1][]
  {\lstset{language=Python,numbers=right,#1}}{}

\newcommand\cpp[1]{\lstinline[language=C++]{#1}}
\newcommand\py[1]{\lstinline[language=Python]{#1}}

\newcounter{bla}

\begin{document}

\begin{frontmatter}



\title{TRIQS: A Toolbox for Research on\\ Interacting Quantum Systems}


\author[CEA]{Olivier Parcollet\corref{author}} \ead{olivier.parcollet@cea.fr}
\author[X]{Michel Ferrero} \ead{michel.ferrero@polytechnique.edu}
\author[CEA,X]{Thomas~Ayral} \ead{thomas.ayral@polytechnique.edu}
\author[CEA]{Hartmut~Hafermann} \ead{hartmut.hafermann@cea.fr}
\author[UHH]{Igor~Krivenko} \ead{ikrivenk@physnet.uni-hamburg.de}
\author[LPTMC,CEA]{Laura~Messio} \ead{messio@lptmc.jussieu.fr}
\author[X]{Priyanka~Seth} \ead{priyanka.seth@polytechnique.edu}

\cortext[author] {Corresponding author.\\\textit{E-mail address:} olivier.parcollet@cea.fr}

\address[CEA]{Institut de Physique Th\'eorique (IPhT), CEA, CNRS, 91191 Gif-sur-Yvette, France}
\address[X]{Centre de Physique Th\'eorique, Ecole Polytechnique, CNRS, 91128 Palaiseau Cedex, France}
\address[UHH]{I. Institut f\"ur Theoretische Physik, Uni. Hamburg, Jungiusstra{\ss}e 9, 20355 Hamburg, Germany}
\address[LPTMC]{LPTMC, UMR 7600 CNRS, Universit\'e Pierre et Marie Curie, 75252 Paris, France}

\begin{abstract}

We present the TRIQS library, a Toolbox for Research on Interacting Quantum
Systems. It is an open-source, computational physics library providing a
framework for the quick development of applications in the field of many-body
quantum physics, and in particular, strongly-correlated electronic systems.  It
supplies components to develop codes in a modern, concise and efficient way:
e.g.~Green's function containers, a generic Monte Carlo class, and simple interfaces to
HDF5.  TRIQS is a \verb#C++#/\verb#Python# library that can be used from either
language.  It is distributed under the GNU General Public License (GPLv3).
State-of-the-art applications based on the library, such as modern quantum
many-body solvers and interfaces between density-functional-theory codes and
dynamical mean-field theory (DMFT) codes are distributed along with it.

\end{abstract}

\end{frontmatter}



\noindent {\bf PROGRAM SUMMARY}

\begin{small}
\noindent
{\em Program Title:} TRIQS                                      \\
{\em Project homepage:} http://ipht.cea.fr/triqs                \\
{\em Catalogue identifier:} --                                  \\
{\em Journal Reference:} --                                     \\
{\em Operating system:} Unix, Linux, OSX \\
{\em Programming language:} \verb*#C++#/\verb*#Python#\\
{\em Computers:} 
  any architecture with suitable compilers including PCs and clusters \\
{\em RAM:} Highly problem-dependent \\
{\em Distribution format:} GitHub, downloadable as zip \\
{\em Licensing provisions:} GNU General Public License (GPLv3)\\
{\em Classification:} 4.4, 4.6, 4.8, 4.12, 5, 6.2, 6.5, 7.3, 7.7, 20 
{\em External routines/libraries:} 
  \verb#cmake#, 
  \verb#mpi#, 
  \verb#boost#, 
  \verb#FFTW#, 
  \verb#GMP#, 
  \verb#BLAS#, 
  \verb#LAPACK#, 
  \verb#HDF5#, 
  \verb#NumPy#, 
  \verb#SciPy#, 
  \verb#h5py#, 
  \verb#mpi4py#, 
  \verb#mako#.\\ 
{\em Nature of problem:}\\
Need for a modern programming framework to quickly write simple, 
efficient and higher-level code applicable to the studies of
strongly-correlated electron systems. \\
{\em Solution method:}\\
We present a \verb#C++#/\verb#Python# open-source computational library that
provides high-level abstractions for common objects and various tools in the
field of quantum many-body physics, thus forming a framework for developing
applications.
{\em Running time:} Tests take less than a minute.
Otherwise highly problem dependent (from minutes to several days).

\end{small}


\section{Introduction}
\label{sec:introduction}

In this paper, we present the 1.2 release of the TRIQS project (Toolbox for
Research in Interacting Quantum Systems), a free software library written in
\verb#Python# and \verb#C++# for the implementation of algorithms in quantum many-body
physics. TRIQS is distributed under the GNU General Public License (GPLv3).
  
Strongly-correlated quantum systems are a central challenge for theoretical
condensed matter physics with a wide range of remarkable phenomena such as
metal-insulator transitions, high-temperature superconductivity and magnetism.
In the last two decades, tremendous progress has been made in the field of
algorithms for the quantum many-body problem, both in refining existing
techniques and in developing new systematic approximations and algorithms.
Methods to address the quantum many-body problem include dynamical mean-field
theory (DMFT)~\cite{Georges96,Kotliar06} and its cluster~\cite{Maier05} or
diagrammatic extensions~\cite{Toschi07,Rubtsov08} or the density matrix
renormalization group (DMRG)~\cite{Schollwoeck05}. DMFT methods can also now
be combined with more traditional electronic structure methods such as density
functional theory (DFT) leading to \emph{ab initio} realistic computational
techniques for strongly-correlated materials~\cite{Kotliar06}.
Several collaborative software development efforts
have made some of these theoretical developments largely accessible, e.g.~Refs. \cite{ALPS20, iquist, itensor}.

The purpose of the TRIQS project is to provide a modern framework of basic building
blocks in \verb#C++# and \verb#Python#. This is needed for the rapid
implementation of a broad spectrum of methods. 
Applications range from simple interactive phenomenological analysis in \verb#Python# to high-performance quantum
impurity solvers in \verb#C++#. At this stage, TRIQS focuses primarily on, but is not
limited to, solid-state physics computations, diagrammatic approximations and
methods of the DMFT family (DMFT, clusters and underlying quantum impurity
solvers).

A particular emphasis is placed on the documentation, in particular in
providing short code examples that can be reused immediately (in \verb#Python# and
\verb#C++#). Full documentation of the project is available online:
\url{http://ipht.cea.fr/triqs}.

Several applications have already been built with the TRIQS library, and some of them are publicly distributed.
Let us mention a state-of-the-art implementation of the hybridisation-expansion quantum impurity solver \textsc{cthyb}
(\url{http://ipht.cea.fr/triqs/applications/cthyb/}) and 
the \textsc{dft tools} project which
provides an interface between DMFT and DFT packages such as \textsc{wien2k} for
realistic computations for strongly-correlated materials
(\url{http://ipht.cea.fr/triqs/applications/dft_tools/}).
Since these applications are not part of the library itself and involve a different set of authors,
they will be presented in separate publications.
However, they are distributed along with the TRIQS library under GPL license and are available for download on GitHub (\url{https://github.com/triqs}).

The TRIQS project uses professional code development methods to achieve the best possible quality for the library and the applications,
including:
{\sl i)} {\it version control} using \verb#git#;
{\sl ii)} {\it systematic code review} by the main TRIQS developers;
{\sl iii)} {\it test-driven development}:
features of the library are first designed with a series of test cases. When the implementation is completed, 
they become the non-regression tests executed during the installation process.

This paper is organised as follows: we start in Sec.~\ref{sec:motivation} 
with the main motivations for the project. In Sec.~\ref{sec:structure}, 
we outline the structure of the TRIQS project. Sec.~\ref{sec:citation} summarizes
our citation policy.
In Sec.~\ref{sec:requirements}, we discuss the prerequisites to
efficient usage of TRIQS and Sec.~\ref{sec:portability} describes the
portability of the library. In Sec.~\ref{sec:example}, we provide two
illustrating examples to give a flavour of the possibilities offered by the
library: we show that TRIQS makes it possible to write a DMFT self-consistency
loop in one page of \verb#Python#, and, in another example, how equations can be coded
efficiently in \verb#C++#. In Sec.~\ref{sec:components} we review the most important
library components. In \ref{app:ctint} we present the implementation of a
fully working, MPI-parallelized, modern continuous-time quantum Monte Carlo
algorithm (the so-called CT-INT algorithm~\cite{Rubtsov05,Gull11}) in about 200 lines. This
example illustrates how TRIQS allows one to design a complex, yet short,
readable and extensible code. 


\section{Motivations}
\label{sec:motivation}

The implementation of modern algorithms for quantum many-body systems
raises several practical challenges.

{\it Complexity}: Theoretical methods and algorithms are becoming increasingly
complex (e.g.~quantum Monte Carlo~\cite{Rubtsov05,Werner06, Gull08, Gull11} and dual
boson~\cite{vanLoon2014} methods). They are hence more difficult to implement, debug and
maintain. This is especially true for realistic computations with methods of
the DMFT family, where one has to handle not only the complexity of the
many-body problem but also the various aspects (orbitals, lattices) of real
materials, which usually requires a well-organised team effort.

{\it Versatility/Agility}: Algorithms are changing and improving rapidly,
sometimes by orders of magnitude for some problems~\cite{Laeuchli09}. This can lead
to a possibly quick obsolescence of a given implementation. To address new
physics problems requires regular and substantial modifications 
of existing implementations. Moreover, there are numerous possibilities for new algorithms which need to be tested quickly.

{\it Performance}: Modern algorithms are still quite demanding on resources, e.g.~quantum Monte Carlo methods.
Hence, the performance of the codes is critical and a simple implementation in a
high-level language is usually not sufficient in practice.

{\it Reproducibility}: The central role and the growing complexity of the
algorithms in our field reinforces the need for reproducibility, which is central to any
scientific endeavour. Therefore, the results obtained by a numerical computation
should in principle be published systematically along with the code that
produced them, in order to allow others to reproduce, falsify or improve on
them.  This requires codes to be readable (i.e. written to be
read by other people than their author) and relatively quick to produce.

To address these challenges, one needs readable, clear and simple
implementations with reusable components provided through high-level
abstractions.
We emphasize that this is \emph{not} in contradiction with the requirement of high performance. The combination of a 
higher level of abstraction with high performance is achieved using modern programming techniques (e.g.~generic programming).
The purpose of the TRIQS project is to find and efficiently implement the relevant abstractions, basic components and algorithms for our domain.

\section{Structure}
\label{sec:structure}

\begin{figure}[h!]
\centering
\caption{Structure of the TRIQS project \label{fig:sketch}}
\includegraphics[width=\textwidth]{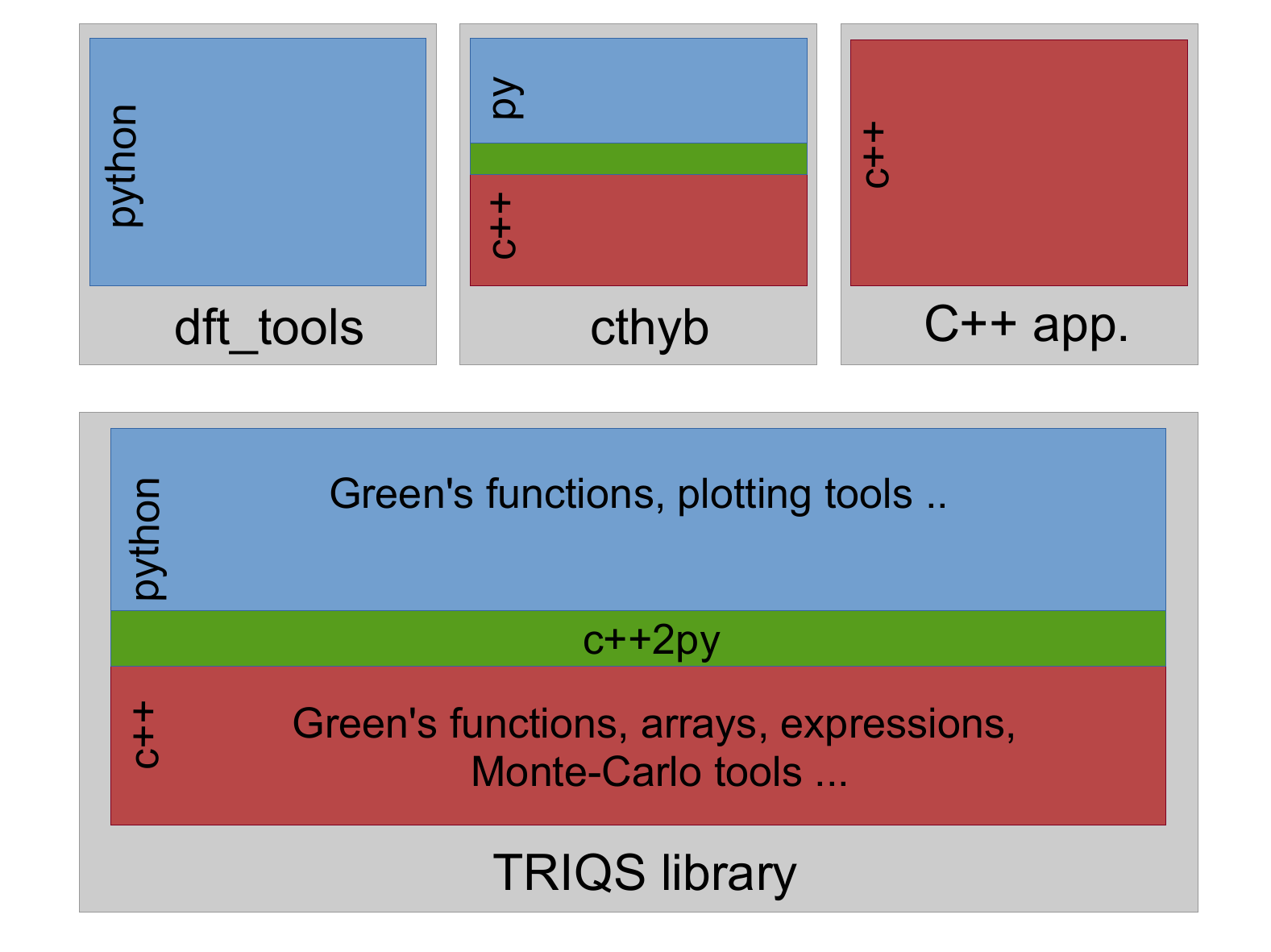}
\end{figure}

The TRIQS framework is depicted in Fig.~\ref{fig:sketch}: the core library is at the root (bottom)
consisting of basic building blocks, which are used in a series of applications (top).
The applications can be in pure python ({\it e.g.} \textsc{dft tools}), in \verb#C++# with a \verb#Python# interface ({\it e.g.} \textsc{cthyb}), 
or even in pure  \verb#C++#.
The subject of this paper is the core library. 

The components of the TRIQS library can be used both in \verb#Python# and in \verb#C++#: \verb#C++#
brings the performance needed for applications where speed is critical (like
many-body solvers) and the type safety of a compiled language. On
the other hand, \verb#Python# is typically used as a higher-level interface for data
analysis, investigation of phenomenological approaches, and tasks related to
reproducibility.  Most objects of the library are written in \verb#C++# and exposed to
\verb#Python# using a specially designed tool described in Sec.~\ref{sec:cpp_python}.  As a result, TRIQS can be used
together with all the modern scientific tools of the \verb#Python# community, in
particular with \verb#IPython# notebooks \cite{IPython} which are recommended for
an optimal interactive usage of the library in \verb#Python#.

\section{Citation policy}
\label{sec:citation}

We kindly request that the present paper be cited in any published work using the TRIQS library directly (e.g.~for data analysis) or indirectly (e.g.~through TRIQS based applications).
In the latter case, this citation should be added to the citations already requested by the application.
This helps the TRIQS developers to better keep track of projects
using the library and provides them guidance for future developments.

\section{Programming Requirements}
\label{sec:requirements}

TRIQS can be used {\sl at different levels of expertise}, starting from basic
\verb#Python# interactive usage to development of cutting-edge mixed \verb#Python#/\verb#C++# high-performance and
massively parallel codes, and in pure \verb#C++#. 

Most objects, in particular the Green's functions, have a rich \verb#Python# interface,
allowing one to easily plot and manipulate them. For example, simple operations
such as value assignment, inversion or output to and input from HDF5 files are all
one-line operations, as shown in the examples below.

At the \verb#C++# level, the required knowledge to make efficient use of the library
is minimised. The systematic usage of {\it modern} \verb#C++# (\verb#C++11# and \verb#C++14#) very often 
lead to simpler syntax than old \verb#C++#.
The library often favours a ``functional style'' programming and the simplest 
possible constructions for the \verb#C++# user.
To fully exploit the capabilities of the library, some understanding of the basic notions of generic programming, such as concepts and templates, is helpful, but not required.
More traditional object-oriented notions of \verb#C++# such as inheritance or
dynamical polymorphism (virtual functions) are not necessary to use the TRIQS library. 

\section{Portability}
\label{sec:portability}

TRIQS is written in {\it modern} \verb#C++#, i.e. using the \verb#C++11# ISO standard.
The motivation for this choice is twofold: first, we encourage the users 
of the library to benefit from the new features of \verb#C++#, in particular those
which produce much simpler code (e.g.~\cpp{auto}, \cpp{for auto} loop or lambdas).
Second, it dramatically reduces the cost of implementing and maintaining the library itself,
since many of the new \verb#C++# features are designed to facilitate the use of the
metaprogramming techniques needed to implement high-level, high-performance libraries.

As a result, TRIQS requires a  \verb#C++11# {\it standard-compliant compiler}.
The documentation provides an updated list of tested compilers.
When it is available, we recommend using a \verb#C++14# compiler for development, in particular 
to get simpler error messages.

At the \verb#Python# level, we use the 2.7 versions of \verb#Python#.  Support for \verb#Python# 3 is planned for later releases.
We use the binary hierarchical data format (HDF5) to guarantee portability of user generated data in binary form.

\section{TRIQS in two examples} 
\label{sec:example}

Here we illustrate the use of the library for two typical tasks encountered in
many-body physics; this should give a flavour of the possibilities offered by
the library.  The first example is a  complete  DMFT computation  implemented in
\verb#Python#, using a continuous-time quantum Monte Carlo solution of the impurity model (which is presented in \ref{app:ctint}). 
The second example illustrates the manipulation of Green's functions in
\verb#C++# within TRIQS.

\subsection{A DMFT computation in one page of IPython}
\label{subsec:dmft}
\begin{figure}[h!]
\centering
\caption{Screenshot of an IPython notebook executing a DMFT loop using the CT-INT solver.\label{fig:pyexample}}
\includegraphics[width=0.99\textwidth]{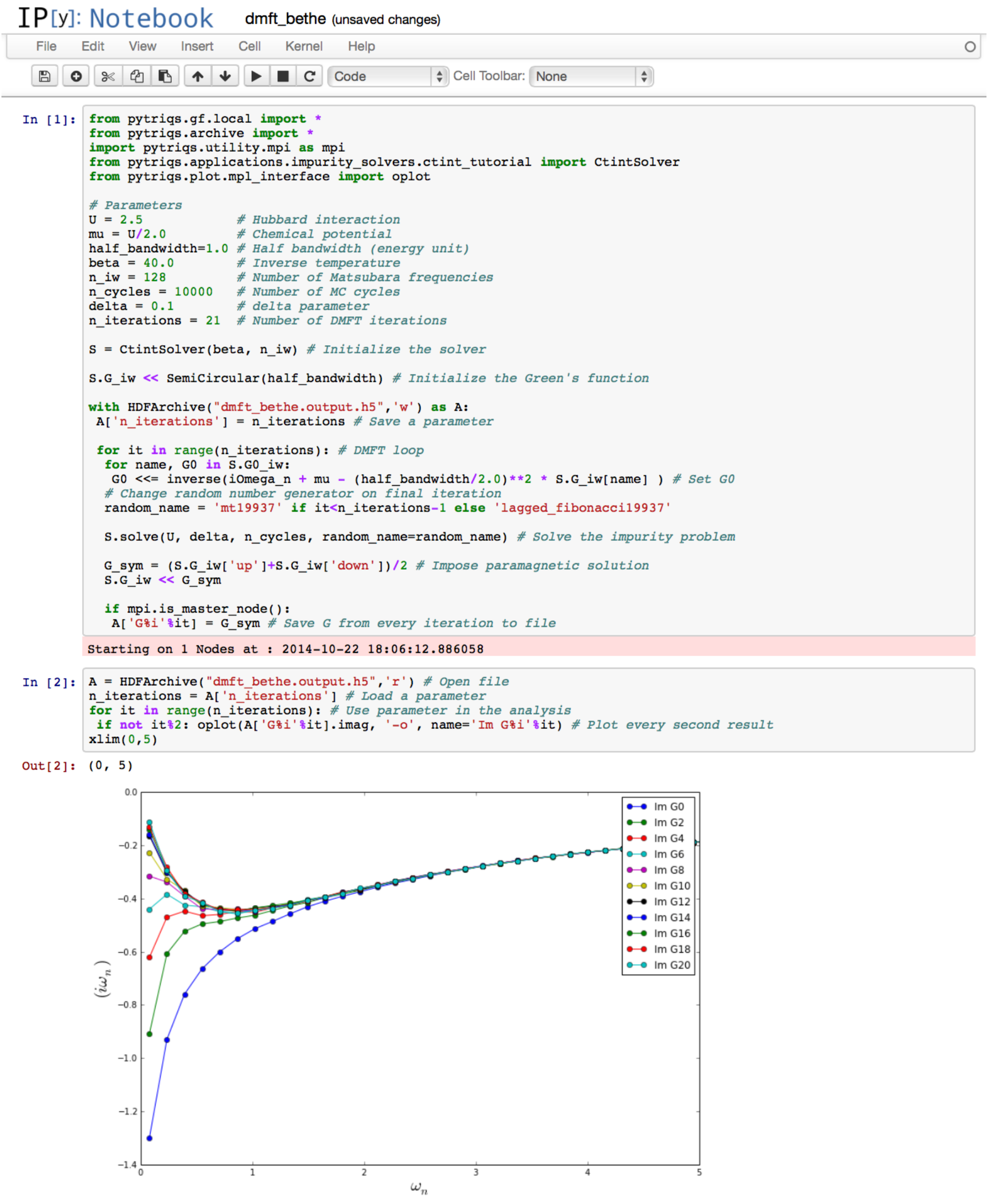}
\end{figure}

This example requires the CT-INT tutorial application to be installed. The installation
procedure is described in Sec.~\ref{sec:installation}.

Fig.~\ref{fig:pyexample} shows a screenshot of an \verb#IPython# notebook implementing
a DMFT self-consistency loop. The essential steps are to load the solver
module, set parameters and an initial guess for the Green's function and to
loop over the DMFT iterations. The solver module, for which performance is
critical, is written in \verb#C++# but used from \verb#Python#. This notebook is available
in the sources of the \verb#ctint_tutorial# application (in the \verb#examples# subdirectory)
along with the corresponding python script that is suitable for parallel execution.

The \verb#Python# framework is \emph{highly flexible}. In this example, we exchange
the random number generator in the final DMFT iteration to consolidate the
result. We could also turn on additional measurements, or implement more
sophisticated stopping criteria for the loop.  Using the \verb#IPython# notebook,
results can be plotted and analysed interactively. As a minimal example, we
load the HDF5 archive from the disk and plot the imaginary part of the Green's
functions in the second cell of the notebook. Parameters used in the
calculation can easily be saved to file and retrieved for data analysis.
On a parallel machine, the first part of the script is executed in Python (without the notebook), 
on multiple cores, while the analysis can still be done in the IPython notebook.

Within this framework, DMFT can readily be explored and practised by
non-experts. The major part of the calculation is of course performed by the
solver module. In this example, we have used the interaction-expansion
continuous-time quantum Monte Carlo (CT-INT) solver. We can easily switch to a more sophisticated
hybridisation-expansion continuous-time quantum Monte Carlo algorithm (CT-HYB)
solver by loading the appropriate solver module instead.  The complete listing
of the \verb#C++# implementation of the solver module is given and explained
in~\ref{app:ctint}, as a more detailed illustration of the library's features.

\subsection{Easy manipulation of Green's functions in C++}

Here, we show how to compute in \verb#C++# a hybridisation function $\Delta(\tau)$  in
imaginary time given the  bare dispersion of a two-dimensional square
lattice with nearest neighbour hopping, at chemical potential $\mu$.
This is a typical task which is usually performed at the
beginning of a DMFT calculation.
The necessary steps are the following:
\begin{align}
G_{0}(i\omega_n) &= \frac{1}{N}\sum_{\kv}  \frac{1}{i\omega_n + \mu -2(\cos k_{x} + \cos k_{y})}
\label{eqn:g0}\\
\Delta(i\omega_n) &= i\omega_n+\mu-G_{0}^{-1}(i\omega_n)
\label{eqn:delta}\\
\Delta(\tau) &= \frac{1}{\beta} \sum_n \Delta(i\omega_n) e^{-i \omega_n \tau} \label{eqn:fft}
\end{align}
The sum over $\mathbf{k}=(k_{x},k_{y})$ is taken over the Brillouin zone, $\omega_n$ is a fermionic Matsubara frequency and $\beta$ is the inverse temperature.
Using the library, these equations are implemented as follows:
\begin{lstlisting}[caption=Computing the hybridisation, label=cppexample,numbers=right]
#include <triqs/gfs.hpp>
using namespace triqs::gfs;
using namespace triqs::lattice;
int main() {

 double beta = 10, mu = 0;
 int n_freq = 100, n_pts = 100;

 // Green's function on Matsubara frequencies, 1x1 matrix-valued.
 auto Delta_iw = gf<imfreq>{{beta, Fermion, n_freq}, {1, 1}};
 auto Gloc     = gf<imfreq>{{beta, Fermion, n_freq}, {1, 1}};

 // Green's function in imaginary time, 1x1 matrix-valued.
 auto Delta_tau = gf<imtime>{{beta, Fermion, 2 * n_freq + 1}, {1, 1}};

 auto bz = brillouin_zone{bravais_lattice{{ {1, 0}, {0, 1} }}};
 auto bz_mesh = regular_bz_mesh{bz, n_pts};

 triqs::clef::placeholder<1> k_;
 triqs::clef::placeholder<2> iw_;

 // The actual equations
 Gloc(iw_) << sum(1/(iw_ + mu - 2*(cos(k_[0]) + cos(k_[1]))), k_ = bz_mesh) 
                /bz_mesh.size();                           // (1)
 Delta_iw(iw_) << iw_ + mu - 1 / Gloc(iw_);                // (2)
 Delta_tau() = inverse_fourier(Delta_iw);                  // (3)

 // Write the hybridization to an HDF5 archive
 auto file = triqs::h5::file("Delta.h5", H5F_ACC_TRUNC);
 h5_write(file, "Delta_tau", Delta_tau);
 h5_write(file, "Delta_iw", Delta_iw);
}
\end{lstlisting}

In the implementation of equations \eqref{eqn:g0} and \eqref{eqn:delta} (lines 23--25), we use a compact syntax for the assignment to the Green's function container provided by the TRIQS library (the CLEF library, Sec.~\ref{sec:clef}).
By definition, this is equivalent to assigning the evaluation of the expression on the right-hand side
to the data points of the ``Green's function''\footnote{Here we refer the Green's function containers and objects representing hybridisation functions or functions with the same signature simply as ``Green's functions''.} \cpp{Delta_iw} on the left and for all Matsubara frequencies in its mesh.
\cpp{iw_} is a \emph{placeholder}, i.e. a dummy variable standing for all points in the Green's function's mesh.

In line 23, the formal expression made of \cpp{iw_} and \cpp{k_} is summed over the values of \cpp{k_}, 
and assigned to the function for each \cpp{iw_}. Note that no copy is made by this statement, 
the computation is inlined by the compiler, as if it was written manually.
This technique is more concise than writing a for-loop on each variable, reduces the risk of errors and simultaneously increases the readability.
Moreover, such techniques come with \emph{no performance penalty} (as indicated by our tests on several standard compilers).

In line, we assign the inverse Fourier transform of $\Delta(\omega)$ to $\Delta(\tau)$ [Eq.~\eqref{eqn:fft}].
Note that the high-frequency expansion is part of the Green's function container and so is {\it automatically} computed in lines 23 and 25 (see Sec.~\ref{sec:gfs} for details). It is used to properly treat the discontinuity in the Fourier transformation in line 26.

Finally, the Green's functions are stored in an HDF5 file with a simple interface,
in a portable manner. The storage conventions are detailed in the reference documentation.

This example is interesting for two reasons: firstly, it shows that the TRIQS library
performs a lot of low-lying operations. There is no need to reimplement them and the user can concentrate on the physics; secondly, it
shows that one can write quite complex operations concisely, which is necessary in order to write readable codes.

\section{Library components}
\label{sec:components}

In this section, we provide an overview of the TRIQS library components.  We
illustrate them either with small examples or in the CT-INT impurity solver
example presented in~\ref{app:ctint}.  The description is neither meant to be
complete nor exhaustive; the online reference documentation
(\url{http://ipht.cea.fr/triqs}) will fill the gaps. 

\subsection{Multidimensional arrays (C++)}
\label{sec:arrays}

TRIQS provides its own multidimensional arrays, with an emphasis on flexibility, performance and the \verb#Python# interface.
It is a fundamental building block for higher-level containers, such as Green's functions.
Listing~\ref{arrayexample} below illustrates some of their features.

\begin{lstlisting}[caption=Array / matrix example, label=arrayexample]
#include <triqs/arrays.hpp>
using namespace triqs::arrays;
int main() {

 auto a = matrix<double>(2, 2);                // Declare a 2x2 matrix of double
 auto b = array<double, 3>(5, 2, 2);           // Declare a 5x2x2 array of double
 auto c = array<double, 2> {{1,2,3}, {4,5,6}}; // 2x3 array, with initialization

 triqs::clef::placeholder<0> i_;
 triqs::clef::placeholder<1> j_;
 triqs::clef::placeholder<2> k_;

 // Assignment of values using CLEF
 a(i_, j_) << i_ + j_;
 b(i_, j_, k_) << i_ * a(k_, j_);

 std::cout << "a = " << a << std::endl; // Printing

 matrix<double> i = inverse(a); // Inverse using LAPACK
 double d = determinant(a);     // Determinant using LAPACK

 auto ac = a;                 // Make a copy (the container is a regular type)
 ac = a * a + 2 * ac;         // Basic operations (uses BLAS for matrix product)
 b(0, range(), range()) = ac; // Assign ac into partial view of b

 // Writing the array into an HDF5 file.
 auto f = triqs::h5::file("a_file.h5", H5F_ACC_TRUNC);
 h5_write(f, "a", a);

 auto m = max_element(abs(b)); // maximum of the absolute value of the array.

 // A more "functional" example: compute the norm sum_{i,j} |A_{ij}|
 auto lambda = [](double r, double x) { return r + std::abs(x); };
 auto norm = fold(lambda)(a, 0);
}

\end{lstlisting}

\begin{shaded}
 
The library provides three types of containers: {\it array} (for multidimensional arrays), {\it matrix} and {\it vector} with 
the following main characteristics: 
\begin{itemize}
\item {\bf Regular-type semantics}: Just like \cpp{std::vector}, these containers have regular-type semantics.
\item {\bf Views}: Each container has a corresponding view type (e.g.~\cpp{array_view}) to e.g.~work on slices and partial views.
\item {\bf CLEF}: The containers are compatible with CLEF (Sec.~\ref{sec:clef} and Listing~\ref{arrayexample}) for fast assignment techniques.
\item {\bf \verb#Python# interface}: These containers can be easily converted to and from \verb#Python# NumPy arrays.
\item {\bf Interface to HDF5}: See Sec.~\ref{sec:h5} and Listing~\ref{arrayexample}.
\item {\bf Arithmetics}: Arithmetic operations are implemented using {expression templates} for optimal performance.
\item {\bf BLAS/LAPACK}: A BLAS/LAPACK interface for matrices and vectors is provided for the most common operations.
\item {\bf STL compatible iterators}: The containers and views can be traversed using such iterators,
      or with simple \cpp{foreach} constructs.
\item \emph{Optionally}, a (slower) debug mode checks for out-of-range operations.

\end{itemize} 

\end{shaded}

\subsection{Green's functions (C++ and Python)}
\label{sec:gfs}

The library provides a special set of containers that allow one to store and
manipulate the various Green's functions used in the quantum many-body problem
and its algorithms. They are defined on meshes for various domains, they are
tensor-, matrix-, or scalar-valued and can be block-diagonal.

Domains currently implemented include real and imaginary frequencies, real and
imaginary times, Legendre polynomials, and Brillouin zones.
Multiple variable Green's functions are also part of the library, but are
restricted to \verb#C++14# mode only and are of alpha quality in release 1.2.
We will not discuss them in this paper.

Green's functions optionally include a description of their high-frequency
behaviour in terms of their moments. Storing this information is important for
several operations (e.g.~Fourier transformation, frequency summation) where
the high-frequency behaviour needs to be treated explicitly. Being part of the
object, the singularity is consistently recomputed in all arithmetic
operations so that the user need not work out the high-frequency
asymptotics. Listing~\ref{gfexample2} illustrates some basic usage of
Green's functions, while a \verb#Python# example has been given above
(Fig.~\ref{fig:pyexample}).

\begin{lstlisting}[caption=Green's function example, label=gfexample2]
#include <triqs/gfs.hpp>
using namespace triqs;
using namespace triqs::gfs;
using namespace triqs::lattice;
int main() {

 double beta = 10;
 int n_freq = 1000;

 clef::placeholder<0> iw_;
 clef::placeholder<1> k_;

 // Construction of a 1x1 matrix-valued fermionic gf on Matsubara frequencies.
 auto g_iw = gf<imfreq>{{beta, Fermion, n_freq}, {1, 1}};

 // Automatic placeholder evaluation
 g_iw(iw_) << 1 / (iw_ + 2);

 // Inverse Fourier transform to imaginary time
 auto g_tau = gf<imtime>{{beta, Fermion, 2 * n_freq + 1}, {1, 1}};
 g_tau() = inverse_fourier(g_iw); // Fills a full view of g_tau with FFT result

 // Create a block Green's function composed of three blocks,
 // labeled a,b,c and made of copies of the g_iw functions.
 auto G_iw = make_block_gf({"a", "b", "c"}, {g_iw, g_iw, g_iw});

 // A multivariable gf: G(k,omega)
 auto bz = brillouin_zone{bravais_lattice{{{1, 0}, {0, 1}}}};
 auto g_k_iw = gf<cartesian_product<brillouin_zone, imfreq>>{
     {{bz, 100}, {beta, Fermion, n_freq}}, {1, 1}};

 g_k_iw(k_, iw_) << 1 / (iw_ - 2 * (cos(k_(0)) + cos(k_(1))) - 1 / (iw_ + 2));

 // Writing the Green's functions into an HDF5 file.
 auto f = h5::file("file_g_k_iw.h5", H5F_ACC_TRUNC);
 h5_write(f, "g_k_iw", g_k_iw);
 h5_write(f, "g_iw", g_iw);
 h5_write(f, "g_tau", g_tau);
 h5_write(f, "block_gf", G_iw);
}

\end{lstlisting}
\begin{shaded}
The main characteristics of Green's functions are:

\begin{itemize}

\item  {\bf Arithmetics}: Just like arrays, Green's functions implement
arithmetic operations using {\bf expression templates}.

\item  {\bf Quick assignment}: The class uses the CLEF component of the TRIQS
library for quick assignment (see Sec.~\ref{sec:clef} and
Listing~\ref{gfexample2}).

\item  {\bf \verb#Python# interface}: The Green's functions are easily shared between
\verb#Python# and \verb#C++#, see Sec.~\ref{sec:cpp_python}, and can thus be used in
conjunction with the \verb#Python# visualisation tools.

\item  {\bf Fourier transforms}: TRIQS provides a simple interface to fast Fourier transforms (FFTW). 
For Green's functions the information about the high-frequency
behaviour is used to avoid numerical instabilities.

\item {\bf Interface to HDF5}.

\end{itemize} 
\end{shaded}

\subsection{Monte Carlo tools (C++)}
\label{sec:mc}

The TRIQS library provides several classes for writing Metropolis-like
(quantum) Monte Carlo algorithms. In addition to some basic analysis tools,
like binning or jackknife, the library mainly contains the \cpp{mc\_generic}
class that implements the Metropolis algorithm (choose a move, try the move,
compute Metropolis ratio, reject or accept, etc.) in terms of
completely generic {\it moves} (configuration updates) and {\it measurements}.

In practice, one just needs to implement the moves and measurements.  The
only requirement is that they must model their respective {\it concepts} \footnote{In
the sense of C++ concepts.}.  For example, the concept of a move is given
by Listing~\ref{lst:mcmove}.  Note in particular that they do not require inheritance
or virtual functions, which makes them particularly simple to use.

\begin{cpplisting}[caption=Concept of a Monte Carlo move, label=lst:mcmove,numbers=none]
struct my_monte_carlo_move {

  // propose a change in the configuration and return the Metropolis ratio
  double attempt(); 
 
  // the move has been accepted: modify configuration
  double accept();  
  
  // the move has been rejected: undo configuration changes
  void reject();  

};
\end{cpplisting}

A concrete usage of the class is shown in the CT-INT solver example~(\ref{app:ctint}).
The class is particularly convenient for complex Monte Carlo algorithms with several
moves: the moves are isolated from the implementation of the Metropolis
algorithm itself and each move can be implemented independently.

The Monte Carlo is of course automatically MPI-enabled. Furthermore, random
number generators can easily be changed dynamically to ensure there is no
subtle correlation effect. 

Listing~\ref{lst:stats} illustrates a basic application of the tools for
statistical analysis on a correlated random series.  Let us assume we have two long
vectors \cpp{V1} and \cpp{V2} storing (possibly correlated) samples of the random variables $X$ and $Y$ and that we wish
to compute estimates of $\langle X\rangle$ and $\langle X\rangle / \langle Y\rangle$, together with the
corresponding error bars. In both cases, the correlation between samples has to be removed using a \emph{binning} procedure. This being done, the first computation is quite straightforward, while the second one further requires a jackknife procedure to remove the bias introduced by the nonlinearity. In TRIQS, all these operations are performed by the following code snippet, using a little library similar to e.g. \verb#ALPS/alea# \cite{ALPS20}:

\begin{cpplisting}[caption=Statistics: error analysis, numbers=none, label=lst:stats,]
 //fill observable with the series 
 observable<double> X, Y;
 for(auto const & x : V1) X << x; //V1: a vector of statistical samples
 for(auto const & y : V2) Y << y; //V2: a vector of statistical samples
  
 std::cout << "<X> is approximately " << average_and_error(X) << std::endl;
 std::cout << "<X>/<Y> is approx.  " << average_and_error(X/Y) << std::endl;

\end{cpplisting}
\cpp{X<<x} fills the observable \cpp{X} (a stack of the samples) with the values \cpp{x} of \cpp{V1}. \cpp{average_and_error(X)} computes an estimate of $\langle X\rangle$ and of the error $\Delta \langle X\rangle$, while \cpp{average_and_error(X/Y)} computes an estimate of $\langle X\rangle/ \langle Y\rangle$ and of $\Delta \langle X\rangle/ \langle Y\rangle$.

\subsection{Determinant manipulations (C++)}
\label{sec:det_manip}

The manipulation of determinants is central to many Monte Carlo approaches to
fermionic problems, see e.g. ~\cite{Rubtsov05,Werner06,Gull08,Gull11}. 
Several cases can be abstracted from the following mathematical problem.
Let us consider a function $F(x,y)$ taking real or complex values (the type of the arguments $x$ and $y$ is arbitrary)
and the square matrix $M$ defined by 
\begin{equation}
  M_{ij} = F(x_i, y_j),
\end{equation}
for two sets of parameters $\{x_i\}$ and $\{y_j\}$ of equal length.
The problem consists in quickly updating $M$ and its inverse $M^{-1}$ 
following successive insertions and removals of one or two lines (labelled by
$x_i$) and columns (labelled by $y_j$) using the Sherman-Morrison and Woodbury
formulas~\cite{Sherman49,Sherman50}.

This generic algorithm is implemented in the TRIQS  \cpp{det_manip} class, 
using BLAS Level 2~\cite{BLAS79,BLAS02} internally. The class provides a simple API, in order 
to make these manipulations as straightforward and efficient as possible. 

For optimal efficiency within a Monte Carlo framework, the modifications to the
matrices can be done in two steps: a first step which only returns the
determinant ratio between the matrix before ($M$) and after the modification ($M'$), i.e. $\xi =
\det M'/\det M$ (which is generally used in the acceptance rate of a Metropolis
move) and a second step which updates the matrix and its inverse. This computationally more expensive
step is usually done only if the Monte Carlo move is accepted.
An example of this class is employed is the CT-INT solver discussed in \ref{app:ctint}.

\subsection{CLEF (C++)} \label{sec:clef}

CLEF (Compile-time Lazy Expressions and Functions) is a component of TRIQS  which
allows one to write expressions with placeholders and functions, and to 
write quick assignments. For example, the following -- quite involved -- equation
\begin{align}
\label{eq:chi0}
  \chi^{0\,\sigma\sigma'}_{\nu\nu'\omega} =
      \beta (g^{0\,\sigma}_{\nu}g^{0\,\sigma'}_{\nu'}\delta_{\omega}
    - g^{0\,\sigma}_{\nu}g^{0\,\sigma}_{\nu+\omega}\delta_{\nu\nu'}\delta_{\sigma\sigma'})
\end{align}
can be coded as quickly as (variables with underscores denote placeholders)
\begin{cpplisting}[numbers=none]
chi0(s_, sp_)(nu_, nup_, om_) << 
      beta * (g[s_](nu_) * g[sp_](nup_) * kronecker(om_)) 
    - beta * (g[s_](nu_) * g[s_](nu_ + om_)
                         * kronecker(nu_, nup_) * kronecker(s_, sp_));
\end{cpplisting}
This writing is clearly much simpler and less error-prone than a more
conventional five-fold nested \cpp{for}-loop. 
At the same time, these expressions are inlined and optimised by the compiler, as if the code were written manually.
The library also automatically optimises the memory traversal (the order of
\cpp{for} loops) for performance based on the actual memory layout of the container
\cpp{chi0}.

The CLEF expressions are very similar to \verb#C++# lambdas, except that their variables
are found by name (the placeholder) instead of a positional argument (in
calling a lambda). This is much more convenient for complex codes.

The precise definition of the automatic assignment is as follows.  Any code of
the form (e.g.~with three placeholders): 
\begin{cpplisting}[numbers=none]
  A(i_,j_,x_) << expression;
\end{cpplisting}
where {\tt expression} is an expression involving placeholders\footnote{For a precise list of what is allowed in expressions, 
the reader is referred to the reference documentation.}
is rewritten by the compiler as follows:
\begin{cpplisting}[numbers=none]
  triqs_auto_assign(
    A, [](auto& i,auto& j,auto& x) { 
             return eval(expression, i_=i, j_=j, x_=x);
        }
  );
\end{cpplisting}
where \cpp{triqs\_auto\_assign} is a free function defined by the container A,
which fills the container with the result of the evaluation of the lambda, and \cpp{eval} evaluates the expression (\cpp{eval} is a function and is part of CLEF). The
precise details of this operation, such as the memory traversal order, are encoded in this
function. The CLEF quick assignment mechanism can therefore easily and efficiently be extended
to any object of the library.
The library provides adaptors to allow standard mathematical functions such as
\cpp{cos} or \cpp{abs} and \cpp{std::vector} to be used in expressions. User-defined
functions and class methods can conveniently be made compatible with the CLEF quick assignment through macros.

\subsection{HDF5 (C++ and Python)}
\label{sec:h5}
HDF5 is  a standard, portable and compact file format, see
\url{http://www.hdfgroup.org}.  Almost all objects in the TRIQS library
(including arrays or Green's functions) can be stored in and retrieved from
HDF5 files, from \verb#C++# and/or \verb#Python#, with a simple and uniform interface. For
example, in \verb#C++#: 
\begin{cpplisting}[numbers=none]
 auto a = array<double, 2> {{1,2,3}, {4,5,6}}; // some data
 {
  auto f = h5::file("data.h5", 'w');           // open the file
  h5_write(f, 'a', a);                         // write to the file
 }                                             // closes the file
\end{cpplisting}
or, the corresponding code in \verb#Python#:
\begin{pylisting}[numbers=none]
 a = numpy.array([[1,2,3],[4,5,6]])
 with HDFArchive("data.h5", 'w') as f:
   f['a'] = a
\end{pylisting}

In \verb#Python#, the HDFArchive behaves in a similarly as a \py{dict}. Therefore, one
can reload a complex object (e.g.~a block Green's function) in a single command in a
script. An example can be seen in Listing~\ref{cppexample}.

An HDF5 file can be seen as a tree whose leaves are ``basic'' objects
(multidimensional rectangular arrays, double, integer, strings, \dots).  More
complex objects are usually decomposed by the library into a subtree of smaller
objects, which are stored in an HDF5 subgroup.  For example, a block diagonal Green's function (of type {\it BlockGf}) is
stored with subgroups containing the Green's functions it is made of; a Green's
function is stored as a subgroup containing the array of data, the mesh, and
possibly the high frequency singularity.  This {\it format}, i.e. the precise
conventions for the names and types of the small objects and the storage order
of the data in the arrays, is described in the reference documentation.
The HDF5 files can be read {\it without the TRIQS library} from \verb#C#,
\verb#C++#, Fortran, \verb#Python# codes, the HDF5 command line tools and with any tool supporting this format. This
enables publishing data and facilitates sharing them across different groups and platforms.
The HDF5 format is indeed widely used, e.g. by the ALPS project \cite{ALPS20}.

Note also that the HDF5 files written from \verb#C++# or \verb#Python# have exactly the same format. Hence one can straightforwardly load some Green's functions in \verb#Python# that have been computed and written using a \verb#C++# code, or vice-versa. 

\subsection{Second-quantized operators (C++ and Python)}
\label{sec:many_body_operator}

The theories of strongly-correlated electron systems often use a language of
second-quantized operators to formulate the problems under consideration.
The model Hamiltonians as well as the observables of interest are routinely
written as polynomials of fermionic operators $c^\dagger$ and $c$.

The TRIQS library implements a \verb#C++# template class \cpp{many_body_operator},
which abstracts the notion of a second-quantized operator. The purpose of this
class is to make expressions for second-quantized operators {\it written in the
\verb#C++# or \verb#Python# code as close as possible to their analytical counterparts}. In order to
pursue this goal, the class implements the standard operator algebra. 
The library stores the expression in normal order, so it performs automatically basic simplications, 
for example when an expression vanishes.
Any operator can be constructed as a polynomial of the elementary operators
carrying an arbitrary number of integer/string indices (defined at compile time).
The coefficients of the polynomials may be real, complex or of a user-defined numeric type in advanced use scenarios.

There is also a \verb#Python# version of the same class (called \py{Operator}),
specialised for the case with real coefficients and the fermionic 
operators with two indices (this particular choice is made for compatibility
with the Green's function component).
Anyone writing a TRIQS-based many-body solver may benefit from this class.
For example, the user of the solver could define a model Hamiltonian in a \verb#Python#
script and subsequently pass it to the solver:

\begin{pylisting}[numbers=none]
from pytriqs.operators.operators import Operator, n, c_dag, c

# Spin operators
Sp = c_dag("up",0)*c("dn",0)        # S_+
Sm = c_dag("dn",0)*c("up",0)        # S_-
Sz = 0.5*(n("up",0) - n("dn",0))    # S_z
S2 = Sz*Sz + (Sp*Sm + Sm*Sp)/2      # S^2

# The Hamiltonian of a half-filled Hubbard atom: four equivalent forms
U = 1.0
H1 = -U/2*(n("up",0) + n("dn",0)) + U*n("up",0)*n("dn",0)
H2 = U*(n("up",0) - 0.5)*(n("dn",0) - 0.5) - U/4
H3 = -2.0*U*Sz*Sz
H4 = -2.0/3.0*U*S2
print H1, '\n', H2, '\n', H3, '\n', H4

# All four forms are indeed equivalent
print (H1-H2).is_zero() and (H2-H3).is_zero() and (H3-H4).is_zero()
\end{pylisting}

\subsection{C++/Python wrapping tool}
\label{sec:cpp_python}

The tool that glues together the \verb#C++# components to \verb#Python# is a crucial 
piece of the TRIQS project. 
Indeed the \verb#C++#/\verb#Python# architecture of the project is very demanding in this aspect:
we need to expose diverse components from \verb#C++# to \verb#Python#. These range from simple functions 
to complex objects with methods, overloaded arithmetic operators, the interface to HDF5, and so on.
The tool must be very flexible, while being as simple as possible to use in the most common cases. 

The TRIQS library proposes such a tool in version 1.2.
From a simple \verb#Python#-written {\it description} of the classes and functions to expose to \verb#Python#, 
it generates the necessary \verb#C# wrapping code to build the \verb#Python# module.
Utilities are also included to actually compile and setup the modules with \verb#cmake#.

In most cases, the process can be fully automatised, using a second tool
based on the \verb#Clang# library, which parses the \verb#C++# code using \verb#libClang# and retrieves the description of the classes and functions
along with their documentation.
As an example, the automatically produced description files for the CT-INT algorithm is provided
in the Appendix.

In more complex cases, some information can be added manually to the class description,
for example the fact that the object forms an algebra over the doubles. 
In such a case, by adding {\it a single line} to the description file, 
the tool automatically generates all the necessary operators for the algebra structure
in \verb#Python# by calling their \verb#C++# counterparts.

\begin{figure}[h!]
\centering
\caption{Using C++ directly within the IPython notebook \label{fig:cppmagic}}
\includegraphics[width=\textwidth]{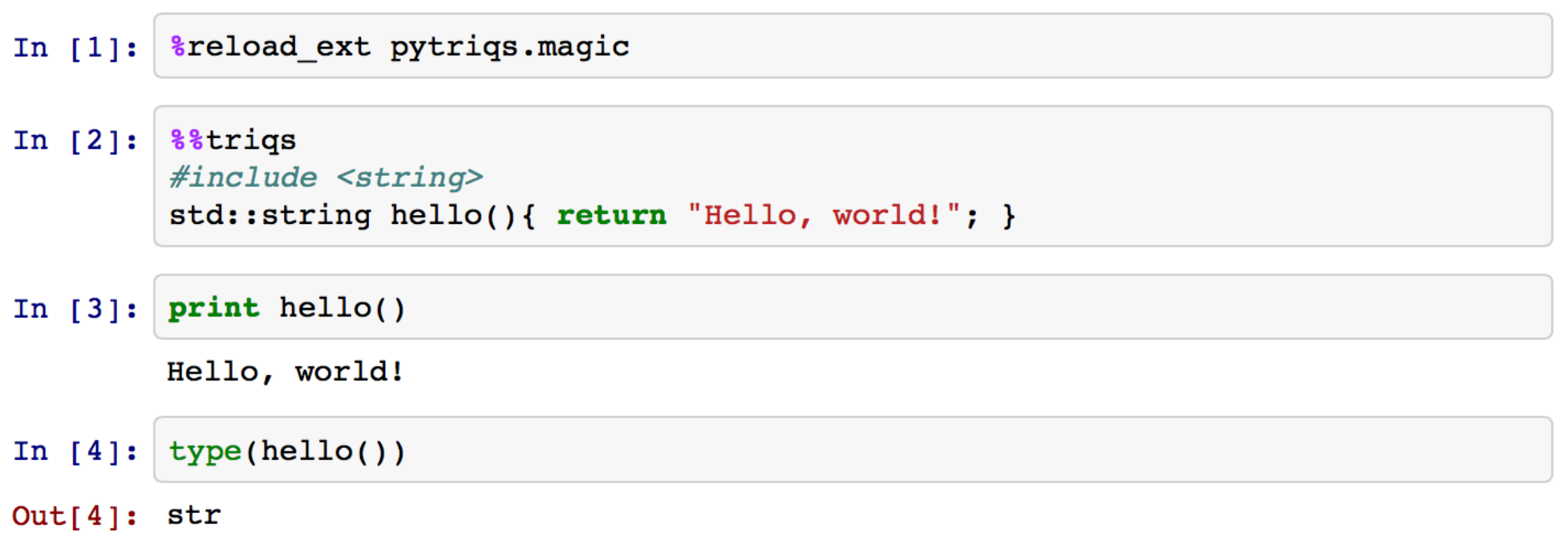}
\end{figure}

As a consequence, this tool also allows the TRIQS user to 
{\it write \verb#C++# code directly within the
\verb#IPython# notebook and use it immediately}, using a so-called ``magic cell command'', in \verb#IPython# terminology.
This is illustrated in Fig.~\ref{fig:cppmagic}.
In this case, the command \py{\%\%triqs} extracts the 
prototype of the \verb#C++# \cpp{hello()} function, writes, compiles and loads the \verb#Python# module 
to be used in the next cell.
TRIQS objects, along with STL containers (e.g.~vector, tuple), 
can be used as function arguments or return values.

Using this feature one can tinker with \verb#C++#
codes directly inside a \verb#Python# environment, without having to set up a \verb#C++#
project.
It is suitable for debugging, quick testing, or executing short \verb#C++# code.
For longer codes, it is better to set up a \verb#Python#/\verb#C++# project along the lines shown for the CT-INT in the Appendix.
Note that this feature is experimental in release 1.2 and currently limited to a single \verb#C++# function per cell
(even though generalisation is quite straightforward).

\section{Getting started}
\label{sec:starting}

Detailed information on installation can be found on the TRIQS website and
current issues and updates are available on GitHub.

\subsection{Obtaining TRIQS}

The TRIQS source code is available publicly and can be obtained by cloning the
repository on the GitHub website \url{https://github.com/TRIQS/triqs}.  As the
TRIQS project is continuously evolving, we recommend that users always
obtain TRIQS from GitHub. Fixes to possible issues are also applied to the
GitHub source.

\subsection{Installation}
\label{sec:installation}

Installing TRIQS is straightforward. We use the \verb#cmake# tool to  
configure, build and test the library.  Assuming that all dependencies have been
installed (refer to the online documentation), the library is simply installed by issuing the following 
commands at the shell prompt:
\begin{verbatim}
$ git clone https://github.com/TRIQS/triqs.git src
$ mkdir build_triqs && cd build_triqs
$ cmake ../src
$ make
$ make test
$ make install
\end{verbatim}
By default, the installation directory \verb#INSTALL_DIR# will be located
inside the build directory. Further installation instructions and help on
installing the dependencies can be found in the online documentation.

\subsection{Usage}
\label{sec:usage}

There are different ways of using TRIQS. In the following, we assume that the
location of the \verb#INSTALL_DIR/bin# folder is in the search path.  We
recommend starting with one of the interactive \verb#IPython# notebook examples
provided with this paper (see below). The interactive \verb#IPython# notebook is started using the
command
\begin{verbatim}
$ ipytriqs_notebook
\end{verbatim}
which will open the browser and allow one to open an existing or a new
notebook. Providing a notebook name as an argument will open the notebook
directly.

The \verb#IPython# example in Fig.~\ref{fig:pyexample} uses the CT-INT solver
of~\ref{app:ctint}, which is shipped as a separate application. Installing
external applications is straightforward. The CT-INT application, for example, is
installed as follows:
\begin{verbatim}
$ git clone https://github.com/TRIQS/ctint_tutorial.git src_ctint
$ mkdir build_ctint && cd build_ctint
$ cmake -DTRIQS_PATH=INSTALL_DIR_ABSOLUTE_PATH ../src_ctint
$ make
$ make test
$ make install
\end{verbatim}
where \verb#INSTALL_DIR_ABSOLUTE_PATH# is the (absolute) path to the TRIQS installation directory.
The application will be installed into the \verb#applications# subdirectory in this TRIQS
installation directory.  Assuming that \verb#INSTALL_DIR_ABSOLUTE_PATH/bin# is in the UNIX search
path, one can then execute the example notebook in Fig.~\ref{fig:pyexample}. To
this end, navigate to the \verb#examples# directory of the
\verb#ctint_tutorial# application sources and issue the following command:
\begin{verbatim}
$ ipytriqs_notebook dmft_bethe.ipynb
\end{verbatim}
This will load the notebook inside a browser. Individual cells can be executed
by pressing \verb#[Shift+ENTER]# (refer to the IPython notebook documentation). The same directory contains a \verb#Python# script
to execute the same DMFT loop from the command line, which is another mode to
use TRIQS that is better suited for long computations on a parallel machine.  It can be executed by typing
\begin{verbatim}
$ pytriqs dmft_bethe.py
\end{verbatim}
or in parallel by running, e.g.,
\begin{verbatim}
$ mpirun -np 4 pytriqs dmft_bethe.py
\end{verbatim}
These commands produce a file \verb#dmft_bethe.output.h5#. To plot the Green's
function from the final iteration, we can launch \verb#ipytriqs# and type:
\begin{verbatim}
$ ipytriqs
...
In [1]: from pytriqs.archive import *

In [2]: from pytriqs.gf.local import *

In [3]: from pytriqs.plot.mpl_interface import oplot, plt

In [4]: A = HDFArchive("dmft_bethe.output.h5","r")

In [5]: oplot(A["G20"].imag, "-o", name="Im G20")

In [6]: plt.show()

\end{verbatim}

As a starting point for developing an external application, we provide a
minimal skeleton application called \verb#hello_world#.  It can be installed in
the same way as the CT-INT solver.
The \verb#C++# examples of this paper, various \verb#IPython# notebooks and the \verb#hello_world# 
are provided in a dedicated GitHub repository \url{https://github.com/TRIQS/tutorials.git}.

\section{Contributing}
\label{sec:contribute}

TRIQS is an open source project 
and we encourage feedback and contributions from the user community to the library and the publication of applications based on it.
Issues should be reported exclusively via the GitHub web site at
\url{https://github.com/TRIQS/triqs/issues}. For contributions, we recommend to use
the {\it pull request} system on the GitHub web site.
Before any major contribution, we recommend to coordinate with the main TRIQS developers.

\section{Summary}
\label{sec:summary}

We have presented the TRIQS library, a Toolbox for Research on Interacting Quantum
Systems. This open-source computational physics library provides a
framework for the quick development of applications in the field of many-body
quantum physics.
Several applications have been built on this library already. They are available
at \url{https://github.com/TRIQS} and will be described in other publications.

\section{Acknowledgements}
\label{sec:acknowledgements}

The TRIQS project is supported by the ERC Grant No. 278472--\emph{MottMetals}.
We acknowledge contributions to the library and feedbacks from M. Aichhorn, A.
Antipov, L. Boehnke, L. Pourovskii, as well as feedback from our user
community. I.K. acknowledges support from Deutsche Forschungsgemeinschaft via
Project SFB 668-A3. P.S. acknowledges support from ERC Grant No.
617196--\emph{CorrelMat}.

\appendix

\section{A sample application: Interaction expansion continuous-time quantum Monte Carlo algorithm}
\label{app:ctint}

In this Appendix, we present the implementation of a simple interaction
expansion continuous-time quantum Monte Carlo algorithm (CT-INT).  We have used
this solver in the DMFT \verb#IPython# example in Fig.~\ref{fig:pyexample}. 
We first briefly recall the formalism of the CT-INT algorithm before discussing the code.

\subsection{Formalism}

We consider the following single-orbital impurity action

\begin{equation}
  S = -\sum_\sigma \int\limits_0^\beta\int\limits_0^\beta d\tau d\tau' \bar d_\sigma(\tau) \tilde{G}^{-1}_{0\sigma} (\tau-\tau')
      d_\sigma(\tau') + \int_0^\beta d\tau \mathcal{H}_\mathrm{int}(\tau),
\end{equation}
whose interaction term is a slightly modified Hubbard term
\begin{equation}
  \mathcal{H}_{int} = \frac{U}{2} \sum_{s=\up,\down}
  \Big(\hat n_\up - \alpha^{s\up} \Big)
  \Big(\hat n_\down - \alpha^{s\down} \Big)
\end{equation}
with
\begin{equation}
  \alpha^{s\sigma} = \frac{1}{2} + (2\delta_{s\sigma} -1)\delta.
\end{equation}
Here $\delta$ is a free small parameter which reduces the sign problem and
$\delta_{s\sigma}$ is a Kronecker symbol.  This rewriting of the interaction
term results in a shift of the chemical potential (absorbed in the  bare
Green's function $\tilde{G}_0$): 
$\tilde{\mu} = \mu - \frac{U}{2}$.  The $\alpha$'s only appear in the
interaction term.
 
The CT-INT algorithm consists in expanding the partition function 
$  Z = \int \mathcal{D}[\bar d,d] e^{-S}$ in powers of $\mathcal{H}_{int}$. 
One obtains:
\begin{multline}
  Z = Z_0 \sum_{k=0}^\infty \frac{(-U)^k}{k!} \int\limits_0^\beta d\tau_1\ldots d\tau_k \frac{1}{2^k} \times\\\times
      \sum_{s_1\ldots s_k=\up,\down}
      \aver{ T_\tau (n_\up (\tau_1) - \alpha^{s_1\up}) \cdots (n_\up (\tau_k) - \alpha^{s_k\up})
      (n_\down (\tau_1) - \alpha^{s_1\down}) \cdots (n_\down (\tau_k) - \alpha^{s_k\down}) }_0.
\end{multline}
where $T_\tau$ is the time ordering operator. 
In the original CT-INT algorithm proposed in Ref.~\cite{Rubtsov05},
$\alpha^{\uparrow \uparrow} = 1 - \alpha^{\uparrow \downarrow} = \alpha$ and
$\alpha^{\downarrow \uparrow} = \alpha^{\downarrow \downarrow} = 0$. This choice can be shown to eliminate
the sign problem for the half-filled single band Anderson model. Here we sum over the indices to make
the formulation slightly more symmetric. It has the advantage that the non-interacting Green's function $\tilde{G}$ 
does not explicitly depend on $\alpha$'s.
Note that in the case of all $\alpha^{s_i \sigma} $ being the same, the sum
over $s_i$ produces a factor $2^k$, which cancels the $2^k$ in the denominator.  The
non-interacting Green's function has no off-diagonal up/down terms, so that the 
average factorises into product of two correlation functions for each spin. Let us furthermore introduce
time ordering by replacing the integrals over the complete time intervals into a
product of time-ordered integrals,
\begin{multline}\label{eq:time-ordering}
  \int\limits_0^\beta  d\tau_1 \hdots d\tau_k
  \aver{\prod_{i=1}^k\prod_\sigma (n_\sigma (\tau_i) - \alpha^{s_i\sigma})}_0
  = k! \int\limits_0^\beta d\tau_1 \int\limits_{0}^{\tau_1} d\tau_2 \hdots \int\limits_0^{\tau_{k-1}} d\tau_k \times\\\times
  \aver{(n_\up (\tau_1) - \alpha^{s_1\up}) \cdots (n_\up (\tau_k) - \alpha^{s_k\up})}_0
  \aver{(n_\down (\tau_1) - \alpha^{s_1\down}) \cdots (n_\down (\tau_k) - \alpha^{s_k\down})}_0.
\end{multline}
Using Wick's theorem and the usual definition for the Green's function
\begin{equation}
  G_{0}^\sigma(\tau) = - \aver{T_\tau d^\pdag_\sigma(\tau) \bar d_\sigma(0)}_0,
\end{equation}
the averages can be represented by determinants. We hence arrive at
\begin{equation}
\label{eqn:Z}
  Z = Z_0 \sum_{k=0}^\infty \int_> d\tau_1\ldots d\tau_k \sum_{s_1\ldots s_k} \frac{(-U)^k}{2^k} \det D^\up_k \det D^\down_k.
\end{equation}
The determinants explicitly read
\begin{equation}
  D^\sigma_k = \left [ \begin{matrix}
  \bar{G}_0^{s_{1}\sigma}(0^-) &  \bar{G}_0^{s_{1}\sigma}(\tau_1 - \tau_2) & \hdots & \bar{G}_0^{s_{1}\sigma}(\tau_1 - \tau_k) \\
    \bar{G}_0^{s_{2}\sigma}(\tau_2 - \tau_1) &  \bar{G}_0^{s_{2}\sigma}(0^-) \hdots & \bar{G}_0^{s_{2}\sigma}(\tau_2 - \tau_k) \\
    \hdots & \hdots & \hdots & \hdots \\
    \bar{G}_0^{s_{k}\sigma}(\tau_k - \tau_1) &  \bar{G}_0^{s_{k}\sigma}(\tau_k - \tau_2)  & \hdots & \bar{G}_0^{s_{k}\sigma}(0^-)
  \end{matrix} \right ],
\label{eq:det}
\end{equation}
where we have defined the Green's function $\bar{G}_{0}^{s\sigma}$ as
\begin{equation}
\bar{G}^{s_{1}\sigma}(\tau_{1}-\tau_{2}) = \left\{\begin{array}{cc}
\tilde{G}_0^\sigma(0^-) - \alpha^{s_1 \sigma} & \tau_{1}=\tau_{2}\\
\tilde{G}_0^\sigma(\tau_{1}-\tau_{2}) & \tau_{1}\neq \tau_{2}\\
\end{array}\right..
\end{equation}
We can sample the partition function \eqref{eqn:Z} by defining a Monte Carlo
configuration as $\mathcal{C}\Let \{ \{\tau_{1},s_{1}\},\dots,
\{\tau_{k},s_{k}\}  \}$ and the Monte Carlo weight of a configuration according
to 
$\omega(\mathcal{C})=\abs{(-U/2)^{k}  \det D^\up_{k} D^\down_{k}}$. The
Metropolis acceptance rate for an insertion of a vertex is
\begin{equation}
  A_{x,y} = \min \Big[ 1 , \frac{-\beta U}{k+1} \frac{\det D^\up_{k+1} D^\down_{k+1}}{\det D^\up_{k} D^\down_{k}} \Big],
\label{eq:insert_ratio}
\end{equation}
while for a removal, it is
\begin{equation}
  A_{x,y} = \min \Big[ 1 , \frac{-k}{\beta U} \frac{\det D^\up_{k-1} D^\down_{k-1}}{\det D^\up_{k} D^\down_{k}} \Big].
\end{equation}
The Green's function can be calculated as
\begin{equation}
G_\sigma(\tau) = - \frac{1}{\beta} \frac{\delta \ln Z}{\delta \Delta^\sigma(-\tau)}.
  \label{eq:G_der_1}
\end{equation}
Carrying out the functional derivative and Fourier transforming yields
\begin{equation}
  G^\sigma(i\omega_n) = \tilde{G}_0^\sigma(i\omega_n) - \frac{1}{\beta} (\tilde{G}_0^\sigma(i\omega_n))^2 
  \sum^\mathrm{MC}_\mathcal{C} \sum_{ij} [ D^\sigma_k ]^{-1}_{ij} e^{i\omega_n(\tau_i - \tau_j)} \text{sign}[\omega(\mathcal{C})]\omega(\mathcal{C}).
\end{equation}
Separating the Monte Carlo weight, we need to accumulate
\begin{align}
M^\sigma(i\omega_n)&\equiv - \frac{1}{Z\beta}
  \sum^\mathrm{MC}_\mathcal{C} \sum_{ij} [ D^\sigma_k ]^{-1}_{ij} e^{i\omega_n(\tau_i - \tau_j)} \times \text{sign}[\omega(\mathcal{C})],
\\
Z &= \sum^\mathrm{MC}_\mathcal{C} \text{sign}[\omega(\mathcal{C})],
\label{eq:m_measure}
\end{align}
From $M$, we can compute the Green's function as follows \cite{Gull08}:
\begin{equation}
\label{eq:compute_g}
G^\sigma(i\omega_n) = \tilde{G}_0^\sigma(i\omega_n) + \tilde{G}_0^\sigma(i\omega_n) M^\sigma(i\omega_n)\tilde{G}_0^\sigma(i\omega_n).
\end{equation}
  
\subsection{Implementation}

As an example of an application of the library, we discuss here the complete
code listing of a fully working, parallelized implementation of the weak-coupling CTQMC
algorithm described above.
How this code can be used in an actual computation is illustrated in the DMFT example of Sec.~\ref{subsec:dmft}.
Through the use of the various components of the library, including
\cpp{gf}, \cpp{mc_tools}, \cpp{det_manip} and \cpp{CLEF}, the full
implementation takes about 200 lines; it comes with a \verb#Python# interface.  Note
that this simple implementation can easily be extended: further 
measurements and moves may be added, or it may be generalised to multi-orbital case or to a
retarded interaction.

We divide the code into several listings that we discuss briefly.
The purpose is to give an illustration of the possibilities of the TRIQS
library without entering into all the details. We start with the main header
file (Listing~\ref{ctint_h}) of the code. It mainly defines and provides
access to the Green's functions that are used in the code, in particular in the main
member function \texttt{solve}.

\begin{lstlisting}[caption=CT-INT: the header file, label=ctint_h, numbers=left]
#include <triqs/gfs.hpp>
#include <boost/mpi.hpp>

// ------------ The main class of the solver -----------------------

using namespace triqs::gfs;
enum spin {up, down};

class ctint_solver {

 block_gf<imfreq> g0_iw, g0tilde_iw, g_iw, M_iw;
 block_gf<imtime> g0tilde_tau;
 double double_occ, percent_done_, beta;
 int n_matsubara, n_times_slices;

 public:

 // Accessors of the class
 block_gf_view<imfreq> G0_iw() { return g0_iw; }
 block_gf_view<imtime> G0_tau() { return g0tilde_tau; }
 block_gf_view<imfreq> G_iw() { return g_iw; }

 ctint_solver(double beta_, int n_iw = 1024, int n_tau = 100001);

 // The method that runs the qmc
 void solve(double U, double delta,
            int n_cycles, int length_cycle = 50, int n_warmup_cycles = 5000,
            std::string random_name = "",
            int max_time = -1);
 
};

\end{lstlisting}

Listing~\ref{ctint_config} defines the Monte Carlo configurations through a
simple vector of determinants~\ref{eq:det} (instances of the \cpp{det_manip}
class). They contain all the necessary information to completely determine a
configuration $\mathcal{C}\Let \{ \{\tau_{1},s_{1}\},\dots, \{\tau_{k},s_{k}\}
\}$. The determinants are constructed from a function object
\texttt{g0bar\_tau}, also declared in this listing, that is used to fill the
elements of the matrix~\ref{eq:det}.

\begin{lstlisting}[caption=CT-INT: define the configurations, label=ctint_config, numbers=left, firstline=6, lastline=45]
// --------------- The QMC configuration ----------------

// Argument type of g0bar
struct arg_t {
 double tau;   // The imaginary time
 int s;        // The auxiliary spin
};

// The function that appears in the calculation of the determinant
struct g0bar_tau {
 gf<imtime> const &gt;
 double beta, delta;
 int s;

 double operator()(arg_t const &x, arg_t const &y) const {
  if ((x.tau == y.tau)) { // G_\sigma(0^-) - \alpha(\sigma s)
   return 1.0 + gt[0](0, 0) - (0.5 + (2 * (s == x.s ? 1 : 0) - 1) * delta);
  }
  auto x_y = x.tau - y.tau;
  bool b = (x_y >= 0);
  if (!b) x_y += beta;
  double res = gt[closest_mesh_pt(x_y)](0, 0);
  return (b ? res : -res); // take into account antiperiodicity
 }
};

// The Monte Carlo configuration
struct configuration {
 // M-matrices for up and down
 std::vector<triqs::det_manip::det_manip<g0bar_tau>> Mmatrices;

 int perturbation_order() const { return Mmatrices[up].size(); }

 configuration(block_gf<imtime> &g0tilde_tau, double beta, double delta) {
  // Initialize the M-matrices. 100 is the initial matrix size
  for (auto spin : {up, down})
   Mmatrices.emplace_back(g0bar_tau{g0tilde_tau[spin], beta, delta, spin}, 100);
 }
};
\end{lstlisting}

Now that the configuration is declared, the next step is to define the Monte
Carlo \emph{moves} that are going to act on this configuration. In
Listing~\ref{ctint_moves}, two moves are implemented: the insertion of an
interaction vertex at a random imaginary time and the removal of a randomly
chosen vertex.  They are described by classes that must model the concept of a
Monte Carlo move. In other words they must have the three members \cpp{attempt},
\cpp{accept}, \cpp{reject}. The \cpp{attempt} method tries a modification of
the configuration and returns a Metropolis acceptance ratio (e.g.~for the
insertion this ratio is given by~\ref{eq:insert_ratio}). The Monte Carlo class
will use this ratio to decide wether to accept or reject the proposed configuration and
then call \cpp{accept} or \cpp{reject} accordingly. Note that for efficiency reasons 
the update of the determinants is done in two steps: in the \cpp{attempt}
method only the ratio of the new to the old determinant is computed (via
\cpp{try_insert}). The actual update of the full inverse matrix is 
performed only if the move is accepted (see the \cpp{complete_operation} call in
\cpp{accept}).

\begin{lstlisting}[caption=CT-INT: define the moves, label=ctint_moves, numbers=left, firstline=46, lastline=93]
// ------------ QMC move : inserting a vertex ------------------

struct move_insert {
 configuration *config;
 triqs::mc_tools::random_generator &rng;
 double beta, U;

 double attempt() { // Insert an interaction vertex at time tau with aux spin s
  double tau = rng(beta);
  int s = rng(2);
  auto k = config->perturbation_order();
  auto det_ratio = config->Mmatrices[up].try_insert(k, k, {tau, s}, {tau, s}) *
                   config->Mmatrices[down].try_insert(k, k, {tau, s}, {tau, s});
  return -beta * U / (k + 1) * det_ratio; // The Metropolis ratio
 }

 double accept() {
  for (auto &d : config->Mmatrices) d.complete_operation(); // Finish insertion
  return 1.0;
 }

 void reject() {}
};

// ------------ QMC move : deleting a vertex ------------------

struct move_remove {
 configuration *config;
 triqs::mc_tools::random_generator &rng;
 double beta, U;

 double attempt() {
  auto k = config->perturbation_order();
  if (k <= 0) return 0; // Config is empty, trying to remove makes no sense
  int p = rng(k);       // Choose one of the operators for removal
  auto det_ratio = config->Mmatrices[up].try_remove(p, p) *
                   config->Mmatrices[down].try_remove(p, p);
  return -k / (beta * U) * det_ratio; // The Metropolis ratio
 }

 double accept() {
  for (auto &d : config->Mmatrices) d.complete_operation();
  return 1.0;
 }

 void reject() {} // Nothing to do
};
\end{lstlisting}

The measurement of the Green's function is shown in
Listing~\ref{ctint_measures}. It is a simple transcription of
Eq.~\ref{eq:m_measure}. Again, the measurements are described by classes that
obey the concept of a Monte Carlo measurement: they have a method \cpp{accumulate}
which is called during the Monte Carlo chain and accumulates data, and a
\cpp{collect_results} method that is called at the very end of the calculation.
Typically the \cpp{collect_results} MPI-reduces the results from several cores
in a parallelized calculation.
Note that \verb#std14::plus# in lines 34 and 35 is the \verb#C++14# version of \verb#std::plus#, which does not require a type,
and which is provided by TRIQS for backward compatibility to \verb#C++11#.

\begin{lstlisting}[caption=CT-INT: define the measures, label=ctint_measures, numbers=left, firstline=94, lastline=132]
//  -------------- QMC measurement ----------------

struct measure_M {

 configuration const *config; // Pointer to the MC configuration
 block_gf<imfreq> &Mw;        // reference to M-matrix
 double beta, Z = 0;

 measure_M(configuration const *config_, block_gf<imfreq> &Mw_, double beta_)
    : config(config_), Mw(Mw_), beta(beta_) {
  Mw() = 0;
 }

 void accumulate(double sign) {
  Z += sign;

  for (auto spin : {up, down}) {

   // A lambda to measure the M-matrix in frequency
   auto lambda = [this, spin, sign](arg_t const &x, arg_t const &y, double M) {
    auto coeff = std::exp(-1_j * M_PI * (x.tau - y.tau) / beta);
    auto fact = coeff * coeff;
    for (auto const &om : this->Mw[spin].mesh()) {
     this->Mw[spin][om](0, 0) += sign * M * coeff;
     coeff *= fact;
    }
   };

   foreach(config->Mmatrices[spin], lambda);
  }
 }

 void collect_results(boost::mpi::communicator const &c) {
  boost::mpi::all_reduce(c, Mw, Mw, std14::plus<>());
  boost::mpi::all_reduce(c, Z, Z, std14::plus<>());
  Mw = Mw / (-Z * beta);
 }
};
\end{lstlisting}

The above components are put together in the main solver body shown in
Listing~\ref{ctint_solve}. The first part is the constructor that only defines
the dimension of the Green's functions. The second part is the \cpp{solve}
method that actually runs the Monte Carlo simulation. It first constructs the
Fourier transform $\tilde{G}_0(\tau)$ of the non-interacting Green's function
given by the user (it is stored in \cpp{g0_iw}). Then a Monte Carlo
simulation is created by adding the relevant moves and measures. This is done
via the \cpp{add_move} and \cpp{add_measure} methods. Note that both the moves
and the measurements are constructed with a reference to the Monte Carlo
configuration \cpp{config}. The simulation is launched with \cpp{start} and 
final results are collected at the end of the simulation with \cpp{collect_results}.
In line 50, we finally compute the actual Green's function through a compact CLEF
expression that implements \eqref{eq:m_measure}.

\begin{lstlisting}[caption=CT-INT: the main solver body, label=ctint_solve, numbers=left, firstline=133]
// ------------ The main class of the solver ------------------------

ctint_solver::ctint_solver(double beta_, int n_iw, int n_tau) : beta(beta_) {

 g0_iw =
     make_block_gf({"up", "down"}, gf<imfreq>{{beta, Fermion, n_iw}, {1, 1}});
 g0tilde_tau =
     make_block_gf({"up", "down"}, gf<imtime>{{beta, Fermion, n_tau}, {1, 1}});
 g0tilde_iw = g0_iw;
 g_iw = g0_iw;
 M_iw = g0_iw;
}

// The method that runs the qmc
void ctint_solver::solve(double U, double delta, int n_cycles, int length_cycle,
                         int n_warmup_cycles, std::string random_name,
                         int max_time) {

 boost::mpi::communicator world;
 triqs::clef::placeholder<0> spin_;
 triqs::clef::placeholder<1> om_;

 for (auto spin : {up, down}) { // Apply shift to g0_iw and Fourier transform
  g0tilde_iw[spin](om_) << 1.0 / (1.0 / g0_iw[spin](om_) - U / 2);
  g0tilde_tau()[spin] = triqs::gfs::inverse_fourier(g0tilde_iw[spin]);
 }

 // Rank-specific variables
 int verbosity = (world.rank() == 0 ? 3 : 0);
 int random_seed = 34788 + 928374 * world.rank();

 // Construct a Monte Carlo loop
 triqs::mc_tools::mc_generic<double> CTQMC(n_cycles, length_cycle,
                                           n_warmup_cycles, random_name,
                                           random_seed, verbosity);

 // Prepare the configuration
 auto config = configuration{g0tilde_tau, beta, delta};

 // Register moves and measurements
 CTQMC.add_move(move_insert{&config, CTQMC.rng(), beta, U}, "insertion");
 CTQMC.add_move(move_remove{&config, CTQMC.rng(), beta, U}, "removal");
 CTQMC.add_measure(measure_M{&config, M_iw, beta}, "M measurement");

 // Run and collect results
 CTQMC.start(1.0, triqs::utility::clock_callback(max_time));
 CTQMC.collect_results(world);

 // Compute the Green function from Mw
 g_iw[spin_](om_) << g0tilde_iw[spin_](om_) + g0tilde_iw[spin_](om_) *
                                                  M_iw[spin_](om_) *
                                                  g0tilde_iw[spin_](om_);
}
\end{lstlisting}

The listings above give a complete implementation of the CT-INT algorithm in
\verb#C++#: the \cpp{ctint_solver} class is ready to be used from within other \verb#C++# programs.
It is however convenient to control calculations on the \verb#Python# level. As
discussed above, TRIQS provides a tool to easily \emph{expose} \verb#C++# to \verb#Python#.
Starting from a descriptor written in \verb#Python# (shown in
Listing~\ref{ctint_wrapper}), it automatically generates a C wrapping code that
constructs the \verb#Python# modules for the \cpp{ctint_solver}. The descriptor 
basically lists the \verb#C++# elements that need to be exposed to
\verb#Python#. In most cases, this descriptor can be generated automatically
by a small analysing tool provided with TRIQS. Here the script~\ref{ctint_wrapper}
has been generated automatically using this tool.

\begin{lstlisting}[caption=CT-INT: Python wrapper descriptor, label=ctint_wrapper, numbers=left, language=Python, breaklines=true]
# Generated automatically using the command :
# c++2py.py ../c++/ctint.hpp -p -m pytriqs.applications.impurity_solvers.ctint_tutorial -o ctint_tutorial
from wrap_generator import *

# The module
module = module_(full_name = "pytriqs.applications.impurity_solvers.ctint_tutorial", doc = "")

# All the triqs C++/Python modules
module.use_module('gf')

# Add here all includes beyond what is automatically included by the triqs modules
module.add_include("../c++/ctint.hpp")

# Add here anything to add in the C++ code at the start, e.g. namespace using
module.add_preamble("""
using namespace triqs::gfs;
""")

# The class ctint_solver
c = class_(
        py_type = "CtintSolver",  # name of the python class
        c_type = "ctint_solver",   # name of the C++ class
)

c.add_constructor("""(double beta_, int n_iw = 1024, int n_tau = 100001)""",
                  doc = """ """)

c.add_method("""void solve (double U, double delta, int n_cycles, int length_cycle = 50, int n_warmup_cycles = 5000, std::string random_name = "", int max_time = -1)""",
             doc = """ """)

c.add_property(name = "G0_iw",
               getter = cfunction("block_gf_view<imfreq> G0_iw ()"),
               doc = """ """)

c.add_property(name = "G0_tau",
               getter = cfunction("block_gf_view<imtime> G0_tau ()"),
               doc = """ """)

c.add_property(name = "G_iw",
               getter = cfunction("block_gf_view<imfreq> G_iw ()"),
               doc = """ """)

module.add_class(c)

module.generate_code()
\end{lstlisting}

After the code has been compiled and installed a new \verb#Python# module is available
in \py{pytriqs.applications.impurity_solvers.ctint_tutorial}. The solver can
then be used as illustrated in Fig.~\ref{fig:pyexample}. As this example
shows, \verb#C++# members like \cpp{g0_iw} can directly be initialised from a \verb#Python#
script and the \cpp{solve} method is also accessible. Controlling the solver, or any other \verb#C++# code 
directly from \verb#Python# makes it very easy to change parameters, plot results, build flexible control structures around it, etc.,
without the need to recompile the codes.

\bibliographystyle{elsarticle-num}
\bibliography{refs.bib}

\end{document}